\documentstyle[11pt,paspconf,epsf]{article}

\begin{document}

\title{Advection-dominated Accretion in the Galactic Center and the Appearance
       of Sgr A$^*$}

\author{Thomas Beckert and Wolfgang J. Duschl\altaffilmark{1}}
\affil{Institut f\"ur Theoretische Astrophysik, Universit\"at Heidelberg,
Tiergartenstr. 15, 69121 Heidelberg, Germany}


\altaffiltext{1}{also at: MPI f\"ur Radioastronomie,
   Auf dem H\"ugel 69, 53121 Bonn, Germany}


\begin{abstract}
We present a modified description of advection-dominated accretion flows
with a geometrical viscosity. A simplified treatment of the dynamics of the
flow in the inner relativistic part is suggested and results are compared with
the galactic center source Sgr A$^*$.
\end{abstract}

\index{Source!Sgr A$^*$}
\index{accretion}
\index{accretion!advection-dominated}

\section{Introduction}

The existence of a massive black hole in the Galactic Center is a keystone
in the understanding of stellar dynamics in the inner parsec
of the galaxy (see the contributions of Ghez and Eckart in this volume)
as it is a test case for the theory of accretion flows in active and non-active
galactic nucleii (Mezger, Duschl \& Zylka, 1996). Standard thin accretion disks
fail to explain the luminosity and spectrum of the radio source
Sgr A$^*$ and several models for the accretion and emission processes have
been suggested. Optically thick synchrotron emission from a jet-disk system
(Falcke et al., 1996) or from an advection-dominated accretion flow (ADAF)
were investigated (Narayan et al., 1998), while optically thin emission by
highly relativistic particles with a quasi-monoenergetic distribution
(e.g. Beckert \& Duschl 1997) is also a possible
mechanism. Mahadevan (1998, and this volume) propose secondary
electrons from pion decay in ADAF's as the synchrotron emitting particles and
showed a way to connect the accretion flow as an energy source with the
relativistic electrons.
  We present a simplified relativistic treatment of advection-dominated
accretion in the galactic center and propose a switch between
standard and ADAF accretion
to distinguish AGN's and sub-luminous galactic centers.

\section{Dynamics of Advection-dominated Accretion into Black Holes}
Advection dominated accretion flows are geometrically thick and a
description in spherical coordinates is appropriate.
In order to aviod complications due to frame
dragging by a rotating black hole we use the Schwarzschild metric and
hope to see also the characteristics of accretion into a slowly rotating hole.
The effects of strong gravity and the existence of a horizon are correctly
described in this way.

\subsection{Energy Advection and Projectors}
The dynamics of an accretion flow into a black hole is governed by
energy and momentum conservation. Energy losses due to radiation have to be calculated
and applied seperately. Angular momentum is transported into the black hole and
to larger radii than those under consideration so a momentum flux
through the boudaries of the disk occurs.
We will neglect magnetic torques in our treatment and consider only a
contribution of the magentic pressure in the flow to the total pressure
by scaling it as $\beta = p_{\rm Gas}/p_{\rm Mag}$.
The dynamics can then be derived from the divergence of the
stress-energy tensor
\begin{equation} \label{fund}
 T^{ij}_{\; ; j} = 0 \quad {\rm with} \quad
   T^{ij} = \tilde\rho \frac{V^iV^j}{c^2} + p g^{ij}
            -2\mu D^{ij} +c^{-2}\left(Q^iV^j + V^iQ^j\right)
            \quad .
\end{equation}
The density $\tilde\rho$ contains a contribution from the specific
enthalpy of the gas like $ \tilde\rho = \rho\left(1 +\frac{h}{c^2}\right)$
with the rest mass density $\rho$. $V^i$ are the components
of the 4-velocity of the flow, $p$ the isotropic pressure and
$g^{ij}$ the metric tensor.
The viscous part of the stress-energy tensor contains the kinematical
viscosity $\mu = \nu \rho$ and the shear-stresses $D^{ij}$.
We have added a contribution by a possible heat flux $Q^i$ which has to be
orthogonal to the velocity.
The 0-component of the divergence in ($\ref{fund}$) is an
energy equation for the flow, but does not resemble
the second law of thermodynamics. This is formulated for a system at rest
so we have to project the divergence in (\ref{fund}) onto the
4-velocity of the flow. The tool for this is a symmetric projector
\begin{equation}
  P^{ij} = g^{ij} + c^{-2}V^iV^j \qquad P^i_j\left(T^{jk}_{\; ; k}\right) = 0
  \qquad .
\end{equation}
After the projection we get an energy equation like
\begin{equation}\label{Enr}
  \rho T \frac{{\rm D}s}{{\rm D} \tau} = 2\mu D_{ij}D^{ij}
   - Q^i_{\; ;i} - c^{-2}Q_iA^i \qquad .
\end{equation}
Here we have used the substantial derivative $\frac{{\rm D}s}{{\rm D} \tau} =
V^j s_{,j}$ and the acceleration $A^i =\frac{{\rm D}V^i}{{\rm D} \tau}$.
In the following we treat stationary and axisymmetric flows with radial infall.

\subsection{Isothermal Poloidal Stratification}
In optically thin plasma the heat produced locally by viscous friction
will be either maintained as internal energy or radiated away. Diffusive
transport by radiation will not occur. Instead of a radiative transport
we expect turbulence
driven by the magneto-rotational instability (e.g. Balbus \& Hawley, 1998)
to transport heat effectively perpendicular to the flow. So we
approximate the poloidal structure by an isothermal plasma in
hydrostatic equilibrium. The poloidal stratification is given by
\begin{equation} \label{veo}
  \frac{\partial p}{\partial \theta} = \rho(1+h/c^2)\sin\theta\ \cos\theta\
  r^2 \Omega^2 \qquad .
\end{equation}
With the assumed isothermal equation of state we can integrate and obtain
\begin{equation}\label{VERTIS}
  \rho(\theta) = \rho_0 \exp\left[-\frac{\cos^2\theta}{2}\left(\frac{r\Omega}
  {c_s}\right)^2 \left(1 + \frac{\epsilon}{c^2} +
  \left(\frac{c_s}{c}\right)^2\right) \right] \qquad .
\end{equation}
One obtains the criterion for the flow to be a thin disk
as
\begin{equation}
  \frac{c_s^2}{1 + \frac{h}{c^2} } \ll \left(r\Omega\right)^2 \qquad .
\end{equation}
We take $c_s^2 =\frac{\partial p}{\partial \rho} \propto T$ at constant
temperature for an ideal gas. The internal energy and so the enthalpy rises
strongly with temperature when the particles become relativistic and
the disk thickness will shrink in the relativistic inner part close to the
black hole.
\subsection{Momentum transport and Viscosity}
In a stationary state as we are interested in the angular momentum
distribution is determined by diffusive transport due to an effective
turbulent viscosity
\begin{equation} 
  -\frac{\dot{M}}{2\pi}\left(\left(1+\frac{h}{c^2}\right)
  \frac{\partial}{\partial r}(r^2\Omega) +
  \frac{r^2 \Omega}{\Sigma c^2}\frac{\partial P}{\partial r}\right)  =
  \frac{\partial}{\partial r}\left( \nu \Sigma r^3
  \frac{\partial}{\partial r} \Omega \right) \qquad .
\end{equation}   
For the viscosity we take a parametrisation
\begin{equation} 
  \nu = \alpha (r-r_s) v_\phi \qquad \alpha \approx {\cal R}_c^{-1}
  \approx 10^{-1\ldots -3}
\end{equation}   
proposed by Duschl, Strittmater \& Biermann which is discussed by R. Auer
(this volume \pageref{RAuer}). For the self-similiar
solution of Narayan \& Yi (1994) this viscosity has the same radial dependence
as the standard form by Shakura \& Sunyaev (1973) . Whenever the ADAF's deviate
from the power-law behavior the viscosity is not coupled to the thermal state
of the flow but rather to the dynamical structure\footnote {The length
scale over which angular momentum is transported
by eddie-exchange is limited by the distance to the horizon at $r_s$.}.
The parameter $\alpha$
is proportional to the inverse of the critical Reynolds number ${\cal R}_c$
in the flow.
MHD-instabilities may produce turbulence at rather small ${\cal R}$, so we
take $\alpha = 0.1$.
The radial momentum balance
\begin{eqnarray} 
  & &\left(\frac{Mc^2}{r^2} - \left(1-\frac{3M}{r}\right)r \Omega^2
  + v_r \frac{\partial v_r}{\partial r}  \right)
  \left(1+\frac{h}{c^2}\right)
  + \frac{\left(1-\frac{2M}{r}\right) + \frac{v_r^2}{c^2}}{\Sigma}
  \frac{\partial P}{\partial r} \nonumber \\
  & & =  \frac{4}{3 \Sigma}\left[
  \left(\frac{\partial}{\partial r} \frac{\nu \Sigma}{r}
  \left(\frac{\partial}{\partial r} r v_r \right) \right) - \frac{2 v_r}{r}
  \left(\frac{\partial}{\partial r} \nu \Sigma \right)
  \right]
\end{eqnarray}   
includes the viscous braking in a converging flow and the inertia and pressure
terms. The centrifugal force reverses sign at $r = 3M = 3/2 R_S$.
The length scale $M = G{\cal M}/c^2$ is the gravitational radius with ${\cal M}$
the mass of the black hole.

\subsection{Heating and Cooling of the Flow}
The thermal balance in a stationary state is given by
\begin{equation}   
\label{EEQ}
Q_{{\rm adv}}  = Q_{\rm vis}^+ - Q_{\rm heat} - Q^- \qquad .
\end{equation}   
The flow cools via bremsstrahlung, synchrotron- and inverse
Compton losses ($Q^-$), while the disk is heated by viscous friction
\begin{equation}   
Q^+_{\rm vis} = \nu \Sigma r^2\left[ \frac{4}{3}\,
\left (\frac{\partial }{\partial r}
\left (\frac {v_r}{r}\right )\right )^{2}
+ \left(\partial_r\Omega \right)^2\right]
\end{equation}   
and the divergence of a possible heat flux $Q_{\rm heat}$. The heat flux
in collisionless plasma is approximated by $Q_{\rm heat} = \Phi \Sigma c_s^3$.
In a magnetizied plasma the heat flux is expected to be strongly depressed so
we take $\Phi = 10^{-3}$. The heat generated in the flow is not completely
radiated but is advected with the flow according to
\begin{equation}  
   Q_{\rm adv} =
   \Sigma v_r c_V\left[\frac{\partial T}
   {\partial r}  - (\Gamma_3 -1)\frac{T}{\rho}
   \frac{\partial \rho}{\partial r} \right]  \qquad .
\end{equation}    
The thermodynamic quantities $c_V, \Gamma_3$ are taken from the
relativistic equation of state for an ideal gas.
The thermal balance must be treated for electrons and ions separately
with an exchange of heat between the two species. This is assumed to
occur only by Coulomb collisions but is subject to debate, as energy transfer
due to resonant scattering of electrons by whistler waves can not be
excluded.
\section{Numerical Treatment}
  The vertically averaged dynamical equations for stationary flows
  presented above allow for a self-similiar solution in the Newtonian limit
and a constant advection ratio (Narayan \& Yi, 1994),
so that $Q_{\rm adv} \propto Q^+_{\rm vis}$
replaces the energy equation (\ref{EEQ}). The self-similiarity implies
constant logarithmic gradients for all variables
\begin{equation}  
  \frac{{\rm d}\Omega}{{\rm d}r} = \left( \frac{{\rm d}\ln \Omega}
   {{\rm d}\ln r}\right)  \frac{\Omega}{r}
\end{equation}    
as shown here for the angular velocity.  The differential
equations are decoupled and one obtains a set of algebraic equations, which
can be easyly solved.
We use a logarithmic radial grid and solve the algebraic
equations with the full energy balance at all points.
From the local solutions we recalculate the actual logarithmic gradients
for a next iteration step.
In doing so we avoid inner regularity conditions
at critical (sonic) points. The steep gradients present in the solutions at
3 Schwarzschild radii (Fig. \ref{temp}) lead to a slow convergence and the
solutions shown are not completely relaxed to a stationary state.
The cooling of the flow by radiation is unimportant at very small
accretion rates and the radiation processes have no backreaction on
the dynamics of the flow. We include bremsstrahlung, synchrotron and
inverse Compton-scattered synchrotron radiation in a rough approximation
in the energy equation and derive the spectra from the ADAF-solutions
afterwards.

\section{Mass Accretion in the Galactic Center}
\index{accretion!mass flow rate}
\begin{figure} 
  \plotone{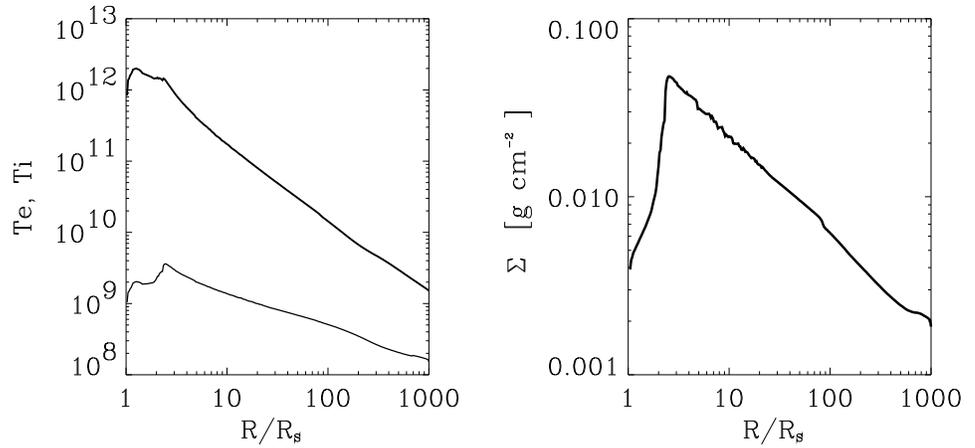}
  \caption{Electron (lower curve) and ion
  temperatures (upper curve) as a function of radius and the surface
  density of an ADAF with a mass flow rate of $10^{-5}$ M$_\odot$/yr.
  Density and temperatures deviate significantly from the self-similiar form
  at the last stable circular orbit at $3 R_S$ where angular momentum does not
  provide a potential barrier anymore. \label{temp}}
\end{figure} 
\begin{figure} 
  \plotone{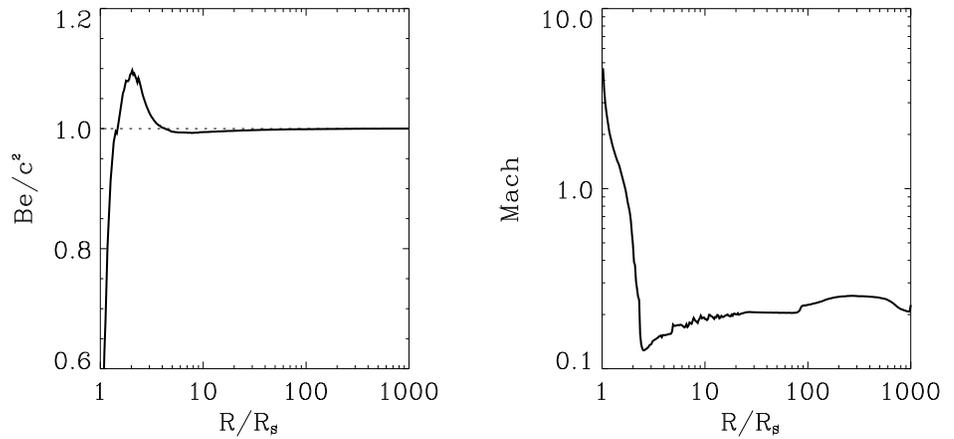}
  \caption{Specific Bernoulli number and radial Mach number of the flow.
  The Bernoulli number includes the rest mass of the particles, so
  a zero-energy flow has a Bernoulli number of 1. The Mach number shows
  a transition from subsonic to supersonic behavior inside the last stable
  circular particle orbit. \label{Bern}}
\end{figure} 
\begin{figure} 
  \plotone{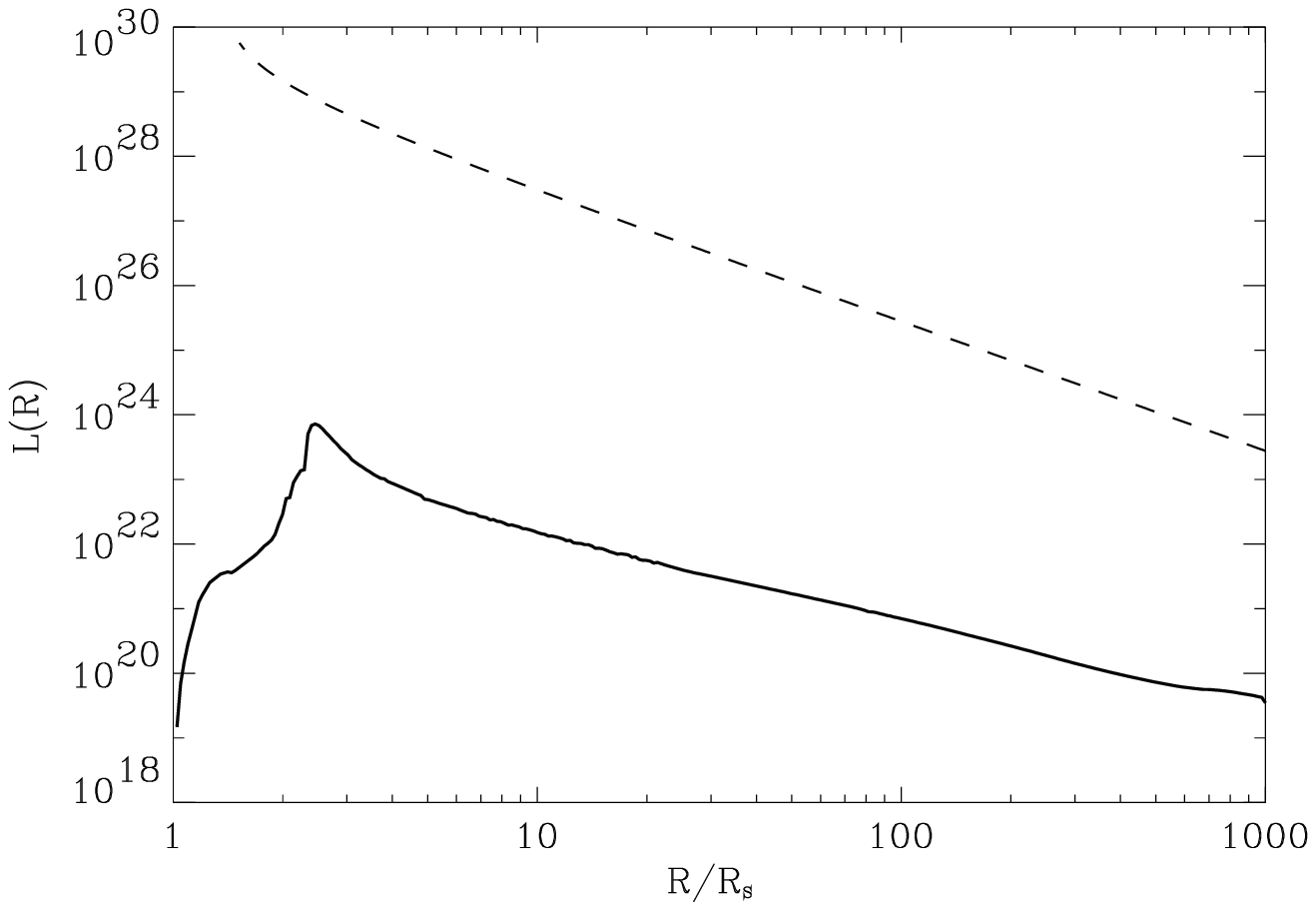}
  \caption{Local luminosity of  ADAF and standard
   thin accretion disk for a mass accretion rate of
   $10^{-5}$ M$_\odot$/yr. The thin disk luminosity (dashed line) is
   considerably larger than the ADAF emission (solid line) and ends at
   $1.5$ black hole radii, where the centrifugal force reverses ist direction
   and is unable to support a stationary thin flow. \label{lum}}
\end{figure} 
The accretion rate necessary to power the galactic center source Sgr A$^*$
is $10^{-9\ldots -10}$ M$_\odot$/yr for a thin accretion disk. But upper
limits from IR-measure\-ments allow only a mass flux of $10^{-11}$ M$_\odot$/yr
or lower. The inefficiency of ADAF-emission processes point to much larger
accretion rates, which are determined from a simultaneous fit of sub-mm and
X-ray measurements. The critical parameters are the mass of the
black hole, here ${\cal M} = 2.6\,10^6$ M$_\odot$, the distance to the
galactic center $D = 8$ kpc and the hydrogen column density for soft X-ray
absorption $N_{\rm H} = 6\,10^{21}$ cm$^{-2}$.
We assume that most radio observations are dominated by optically thin
synchrotron emission from relativistic electrons
and the ADAF contributes significantly in the sub-mm regime only. This minimal
ADAF-model is calculated with a plasma-$\beta = 10$ and is correspondingly
gas pressure dominated. The derived mass accretion rate is
$1\cdot 10^{-5}$ M$_\odot$/yr. Temperatures and surface density of this model
is shown in Fig. \ref{temp}. In contradiction to the analytical solutions and
Narayan \& Yi (1995) we find Bernoulli numbers less than 1, corresponding to
the rest mass at infinity as seen in Fig. \ref{Bern}.
The radiation inefficiency is shown in Fig. \ref{lum}, where the local luminosity
is compared  to the luminosity of a thin disk at the same accretion rate.

\section{Synchrotron and X-ray Spectrum of Sgr A$^*$}
\index{accretion!advection-dominated!emission}
\begin{figure} 
  \plotone{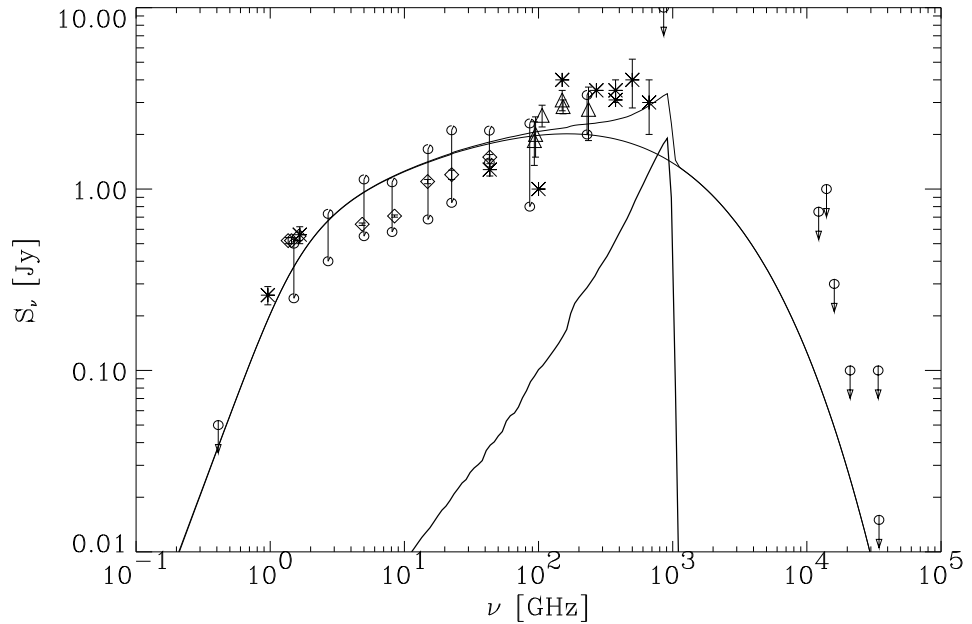}
  \caption{A comparision between minimal ADAF model and radio-IR observations.
  We have combined single measurements with variability ranges marked by
  symbols with circels at both ends. The emission by the ADAF itself peaks at
  $10^3$ GHz. Optically thin synchrotron emission from relativistic
  electrons provide the flux at almost all other frequencies.
  The combined spectrum is also shown and falls a little short for
  the measured flux at 100-1000 GHz. }
\end{figure} 
\begin{figure} 
  \plotone{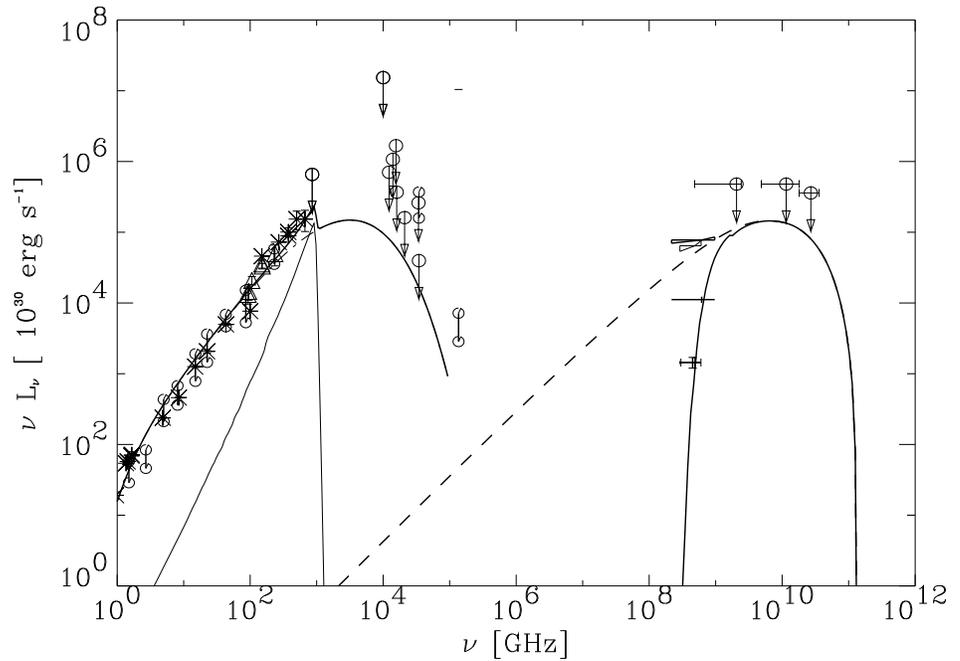}
  \caption{The radio to X-ray emission of Sgr A$^*$ and the minimal
   ADAF-model. ROSAT, {\it Einstein} and GINGA observations are included. Both
   the uncorrected and the corrected fluxes corresponding to a
   column density of $N_{\rm H} = 6\,10^{21}$ cm$^{-2}$ are shown.
   The intrinsic spectrum
   of free-free emission is given by the dashed line. According to our model
   most radio observations see optically thin synchrotron emission,
   but the energy output is
   dominated by the ADAF, both in the sub-mm and X-ray regime. The comptonized
   synchrotron emission emerges at IR-optical frequencies and is not included.
   \label{radio}}
\end{figure} 
The optically thick self-absorbed synchrotron radiation of the ADAF
emerges in the radio and sub-mm regime and rises
steeply to a maximum at $10^3$ GHz. It shows an even steeper cut-off for
larger frequencies seen in Fig. \ref{radio}. Most of the radio observations can not be
explained by the ADAF itself. One possible source for this radiation
are relativistic electrons of $E = 203$ MeV emitting optically thin
synchrotron radiation in a weak magnetic field  $B=2.8$ G.
Assuming a homgeneous source its radius is at least $R = 4.5\,10^{13}$ cm,
much larger than the synchrotron emitting region of the ADAF and thus
providing an extended halo. These relativistic electrons can be
accelerated in magnetic reconnection sheets (Schopper, Lesch \& Birk, 1998)
and maintained in a quasi-monoenergetic distribution (Jauch, this volume)
or they may be secondary electrons from charged pion decays (Mahadevan, 1998).
\section{How to switch on an AGN and how to switch it off again}
\index{acretion!AGN-switch}
\begin{figure} 
  \plottwo{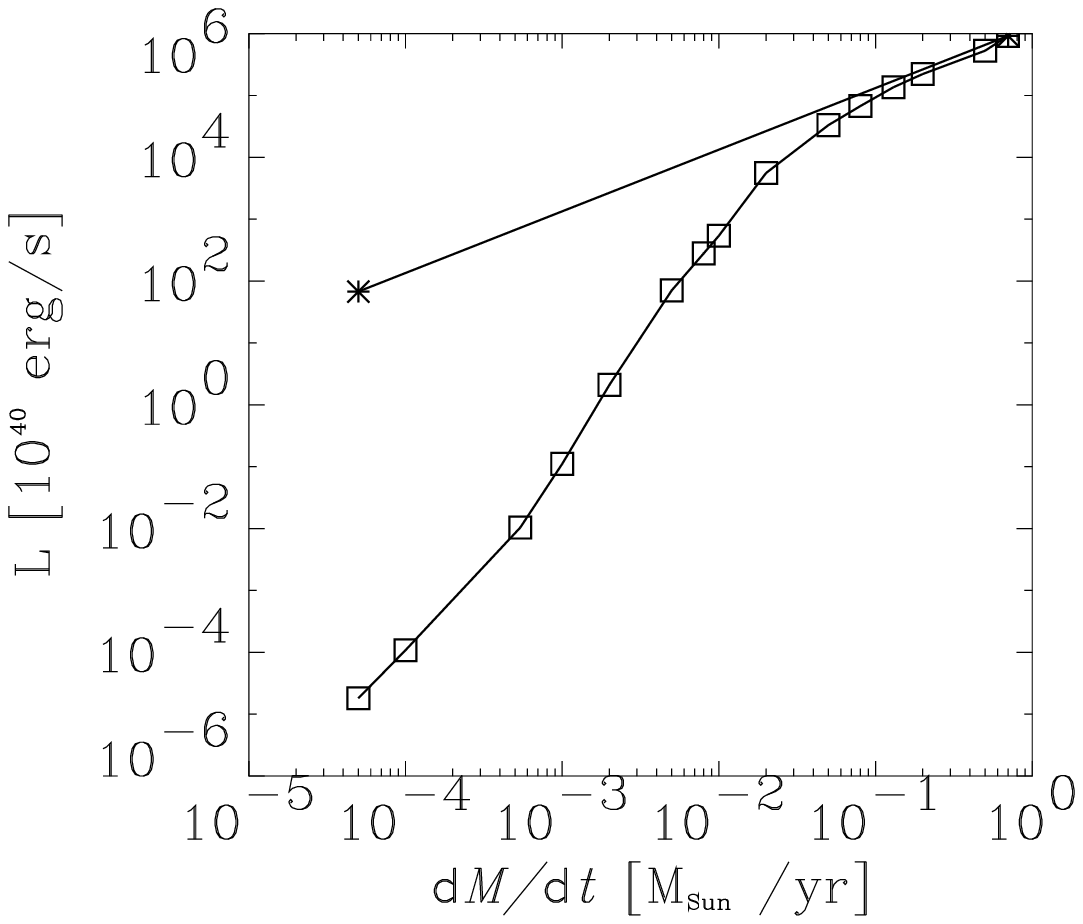}{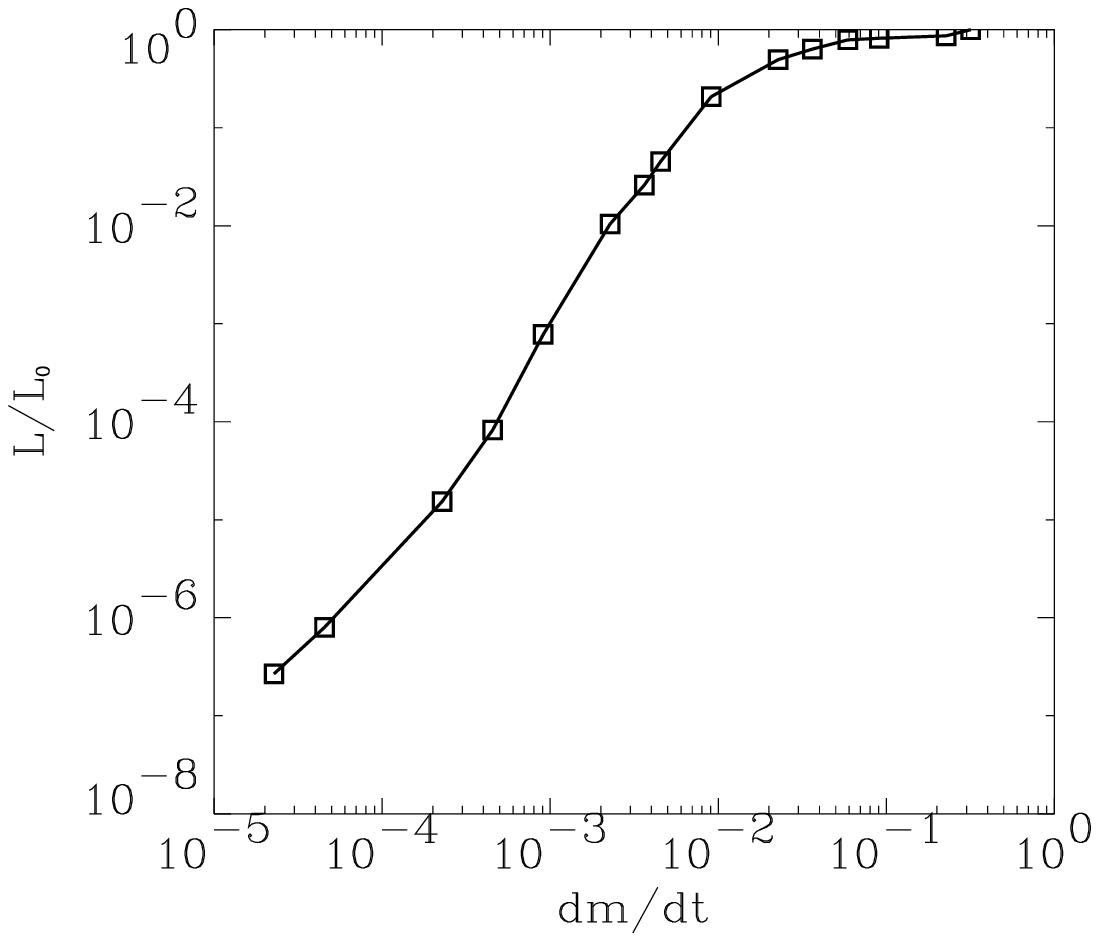}
  \caption{The total luminosity of standard (upper line) and ADAF's as a
  function of mass accretion rate (left plot) for a $10^8$ M$_\odot$
  black hole. Squares mark calculated luminosities of ADAF's in our treatment.
  The relative luminosity of ADAF's versus $\dot{M}$ in units of the
  Eddington accretion rate shows the rapid deviation of ADAF's from the
standard disk luminosity within two decades in the mass flow rate. }
\label{agn-switch}
\end{figure} 
For mass accretion rates below 0.1 $M_{\rm Edd}$ there are two possible
states of accretion into black holes. The standard thin disk
(Frank, King \& Raine, 1992 ), which looses
all the energy produced by viscous friction locally, and
the advection-dominated accretion flow (ADAF) described above. When we compare
the radially integrated luminosity of the two kinds of accretion flows at
the same accretion rate, we find roughly the same luminosity for
a critical mass flow rate of $\alpha^2 \dot{M}_{\rm Edd}$. At smaller rates the
luminosity of ADAF's drop dramatically, while the luminosity of efficiently
radiating flows is proportional to the mass accretion rate $L_0 = \eta \dot{M} c^2$.
The ratio of the luminosities for ADAF's to $L_0$ gives the amount of
specific energy advected into the black hole and the inefficiency of the
heat transfer between electrons and ions.

As both states of accretion are available at the same time in a stationary
flow, there must be a physical reason, which enforces the flow to be
advection-dominated or not. If a mechanism is at work deciding in which
state the flow will be, it provides a switch to turn a luminous accreting black
hole into a quiet sink of matter or the other way round.

Let us, for the moment being, assume that, whenever available, an
accretion disk chooses to operate in the advection dominated
regime, which is indicated by the extremely low luminosity of the
Galactic Center. Then comparatively small changes in the mass flow
rate through the disk may result in dramatic changes of the
efficiency of the accretion process, i.e., the luminosity of the
AGN. This highly non-linear behaviour of the luminosity as a
function of the accretion rate has the potential of acting almost
as a switch between an AGN phase of a galaxy and a non-active
phase whithout having to change the mass flow rate by a similar
amount as the luminosity. In our example of Fig.\
\ref{agn-switch}, at a mass flow rate of $\sim 10^{-4}\,{\rm
M}_\odot$/yr an increase of the rate by two orders of magnitude
results in almost a million times larger a luminosity.


\acknowledgments
This work was supported by the DFG through SFB 328 {\it Evolution of Galaxies}.

%
%


\end{document}